\documentclass[twocolumn,aps,superscriptaddress,prl]{revtex4}
\usepackage{graphicx}
\usepackage{xcolor}
\usepackage{times}
\usepackage{amsmath}
\usepackage{amssymb}
\usepackage{units}
\usepackage{stmaryrd} 
\DeclareGraphicsExtensions{.pdf,.png,.eps,.jpg}

\begin{document}
\preprint{0}

\title{Ultrafast optical control of the electronic properties of $\mathbf{ZrTe_5}$}

\author{G. Manzoni} 
\affiliation{Universit\'a degli Studi di Trieste - Via A. Valerio 2, Trieste 34127, Italy} 

\author{A. Sterzi}
\affiliation{Universit\'a degli Studi di Trieste - Via A. Valerio 2, Trieste 34127, Italy} 

\author{A. Crepaldi} \email[E-mail address: ]{alberto.crepaldi@elettra.eu}
\affiliation{Elettra - Sincrotrone Trieste, Strada Statale 14 km 163.5 Trieste, Italy} 

\author{M. Diego}
\affiliation{Universit\'a degli Studi di Trieste - Via A. Valerio 2, Trieste 34127, Italy} 

\author{F. Cilento} 
\affiliation{Elettra - Sincrotrone Trieste, Strada Statale 14 km 163.5 Trieste, Italy} 

\author{M. Zacchigna}
\affiliation{C.N.R. - I.O.M., Strada Statale 14 km 163.5 Trieste, Italy}

\author{Ph. Bugnon}
\affiliation{Institute of Condensed Matter Physics, Ecole Polytechnique F\'ed\'erale de Lausanne (EPFL), CH-1015 Lausanne, Switzerland}

\author{H. Berger}
\affiliation{Institute of Condensed Matter Physics, Ecole Polytechnique F\'ed\'erale de Lausanne (EPFL), CH-1015 Lausanne, Switzerland}

\author{A. Magrez}
\affiliation{Institute of Condensed Matter Physics, Ecole Polytechnique F\'ed\'erale de Lausanne (EPFL), CH-1015 Lausanne, Switzerland}

\author{M. Grioni}
\affiliation{Institute of Condensed Matter Physics, Ecole Polytechnique F\'ed\'erale de Lausanne (EPFL), CH-1015 Lausanne, Switzerland}

\author{F. Parmigiani}
\affiliation{Universit\'a degli Studi di Trieste - Via A. Valerio 2, Trieste 34127, Italy} 
\affiliation{Elettra - Sincrotrone Trieste, Strada Statale 14 km 163.5 Trieste, Italy} 
\affiliation{International Faculty - University of K\"oln, 50937 K\"oln, Germany} 

\date{\today}

\begin{abstract}

We report on the temperature dependence of the $\mathrm{ZrTe_5}$ electronic properties, studied at equilibrium and out of equilibrium, by means of time and angle resolved photoelectron spectroscopy (tr-ARPES). Our results unveil the dependence of the electronic band structure across the Fermi energy on the sample temperature. This finding is regarded as the dominant mechanism responsible for the anomalous resistivity observed at $\mathrm{T^*} \sim 160$ K along with the change of the charge carrier character from hole-like to electron-like. Having addressed these long-lasting questions, we prove the possibility to control, at the ultrashort time scale, both the binding energy and the quasiparticle lifetime of the valence band. These experimental evidences pave the way for optically controlling the thermo-electric and magneto-electric transport properties of $\mathrm{ZrTe_5}$.

\end{abstract}

\maketitle

Transition metal pentatelluride, $\mathrm{ZrTe_5}$, displays a rich set of unique and exotic transport properties, that make this material an emerging candidate for magnetic and thermoelectric devices \cite{Sambongi_book, Tritt_book}. The complexity of the  $\mathrm{ZrTe_5}$ electronic properties became clear since its first synthesis in 1973 \cite{Furuseth_1973}. The thermopower changes sign from positive to negative when cooling down across $\mathrm{T^*} $ \cite{Jones_1982}, meanwhile the resistivity increases by a factor $\sim 3$ \cite{Skelton_1982}, and the charge carrier switches from holes, at $\mathrm{T}>\mathrm{T^*} $, to electrons, for $\mathrm{T}<\mathrm{T^*} $ \cite{Izumi_1982}. Several models have been proposed to interpret these transport properties, based on charge density wave \cite{Jones_1982}  or polaron \cite{Rubinstein_PRB_1999} formation, but so far not supported by direct experimental evidences \cite{DiSalvo_PRB_1981, Okada_1982}.

The discovery of large magnetoresistance, with both positive \cite{Tritt_PRB_1999} and, more surprisingly, negative sign \cite{Valla_arx_2015}, has proved that the magnetic and electronic transport properties are tightly connected in $\mathrm{ZrTe_5}$. These findings have been recently regarded as a sign of chiral magnetic effect, possible in the case of 3D Dirac semimetal \cite{Valla_arx_2015, Moll_arx_2015, Chen_arx_2015, Chen_arXiv_2_2015}. Furthermore, $\mathrm{ZrTe_5}$ exhibits a superconducting state under pressure \cite{Zhou_arx_2015} while spin helical surface states have also been predicted \cite{Weng_PRX_2014, Hasan_RMP_2010}. All together these properties require further thorough investigations of the physical properties of this material, with particular attention to the interplay between temperature and electronic structure, with the aim to clarify the still unknown origin of the resistivity anomaly.


\begin{figure*}[t!]
  \centering
  \includegraphics[width = 0.8\textwidth]{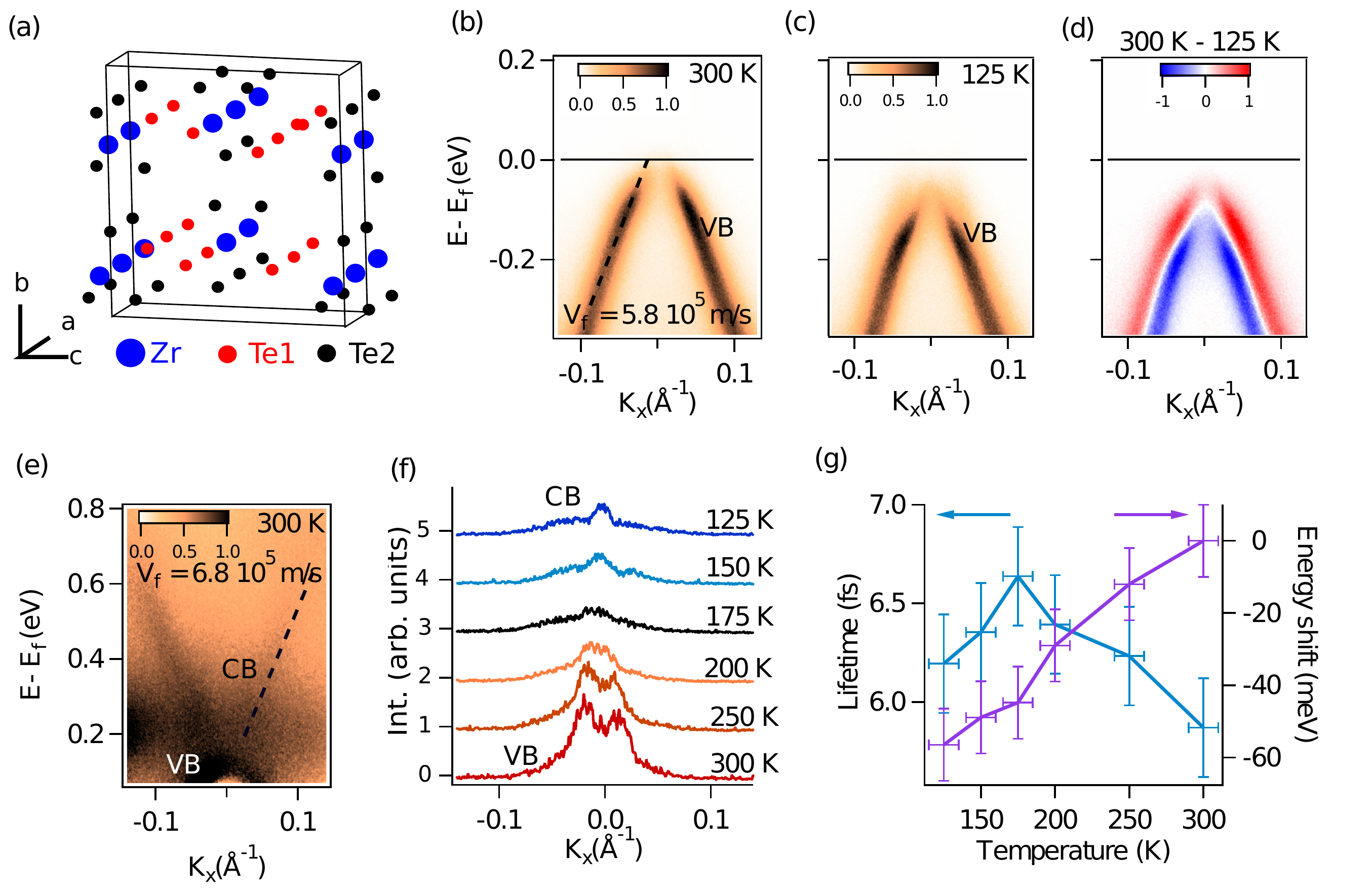}
  \caption[didascalia]{ (color online) (a) Schematic of the $\mathrm{ZrTe_5}$ crystal structure, formed by chains of $\mathrm{ZrTe_3}$ (Zr in blue and Te2 in black), running along the \emph{a} crystallographic direction and linked along the \emph{c} direction by  zig-zag chains formed by Te1 atoms (in red)  \cite{Fje_1986, Weng_PRX_2014}. (b) - (c) Electronic band structure of $\mathrm{ZrTe_5}$ along the chain direction $\mathrm{\Gamma X}$ measured at 300 K and 125 K respectively, using s-polarized light. A linear fit of the valence band (VB) dispersion, dashed line in (b), results in a band velocity equal to $\sim 5.8 \times 10^5$ m/s. (d) difference image obtained by subtracting the low temperature band dispersion from the high temperature. Blue and red indicate the transfer of spectral weight to lower energies when cooling down. (e) Unoccupied conduction band (CB), probed with p-polarized light after optical transfer of charges from the occupied VB. The band velocity, extracted from a linear fit, corresponds to $\sim 6.8 \times 10^5$ m/s. (f) Evolution of the spectral weight in proximity of the Fermi Level $\mathrm{E_F}$. MDCs are integrated 20 meV below $\mathrm{E_F}$ and positively shift along the vertical axis for the different sample temperatures. The spectrum evolves from the double peaks of the VB crossing $\mathrm{E_F}$ at room temperature, to a minimum of intensity at 175 K followed by an increase of intensity which we attribute to the appearance of CB bottom at lower temperature. (g) Evolution of the binding energy shift and the quasiparticle lifetime as extracted from the fit of the MDCs at $-0.12$ eV as a function of the temperature. The increasing lifetime from room temperature to 175 K is signature of metallic behavior. Whereas, at temperature lower than 175 K the lifetime decreases, indicating an increased scattering rate when the VB state lies completely below $\mathrm{E_F}$.
  
    }
  \label{fig:staticARPES}
\end{figure*}


In this letter, by combining angle resolved photoelectron spectroscopy (ARPES) and time resolved ARPES (tr-ARPES) we unveil the evolution of the electronic structure versus temperature in $\mathrm{ZrTe_5}$. Moreover, by studying the out-of-equilibrium band structure we prove the possibility of optically controlling the electronic properties of $\mathrm{ZrTe_5}$. All together our findings indicate the way to manipulate, at the ultrashort time scale, both the thermo-electric and magneto-electric transport properties.

One of the key points of our experiment is the direct imaging of the conduction band (CB). This band, located above the Fermi level ($\mathrm{E_F}$), is transiently populated and detected via non-linear two-photons photoemission (2PPE) process. Temperature dependent experiments reveal a negative energy shift of the $\mathrm{ZrTe_5}$ band structure. The same binding energy shift is observed also by optically exciting the system out-of-equilibrium. The combined knowledge of both the occupied and unoccupied band structure, along with its temporal evolution across $\mathrm{T^*}$, enable us to account for the origin of the anomalous resistivity peak and the charge carrier switch from holes, at $\mathrm{T}>\mathrm{T^*}$, to electron, at $\mathrm{T}<\mathrm{T^*}$.  Having clarified the electronic mechanisms that control the transport properties in  $\mathrm{ZrTe_5}$, we prove that, by means of an ultrafast optical excitation it is possible to directly control the energy position of the band structure and the quasiparticle lifetime in the occupied valence band (VB) states.


The tr-ARPES experiments are performed at the T-ReX Laboratory, Elettra (Trieste, Italy), operating a Ti:Sapphire regenerative amplifier (Coherent RegA 9050) working at 250 kHz repetition rate, whose laser fundamental emission centered at 800 nm ($\sim$1.55 eV) is split into two beams. The p-polarized pump beam excites the $\mathrm{ZrTe_5}$ samples with an impinging fluence $\sim 100~\mu J/cm^2$. Electrons are photoemitted by the probe beam, corresponding to the laser $\mathrm{4^{th}}$ harmonics  ($\sim 6.2$ eV) generated in phase-matched barium borate (BBO) crystals. A $\lambda /2$  waveplate is used for setting the \emph{s} or \emph{p} polarization state for the probe beam. The photoelectrons are collected and analyzed by a SPECS Phoibos 225 hemispherical spectrometer, with energy and angular resolution set to 10 meV and 0.2 $^{\circ}$,  respectively. The overall temporal resolution is equal to 250 fs \cite{Crepaldi_2012}.
High quality  $\mathrm{ZrTe_5}$ single crystal are grown by direct vapour transport technique with iodine  methods \cite{Berger_1983}. The samples are cleaved in ultra high vacuum and mounted on a variable temperature cryostat. The precise temperature of the resistivity anomaly slightly depends on the sample growth conditions \cite{Rubinstein_PRB_1999}, and for the present study it corresponds to $\mathrm{T^* \sim 160}$ K  \cite{Jens_private}.


$\mathrm{ZrTe_5}$ crystallize in the layered orthorombic crystal structure and it belongs to the $CmCm~(D^{17}_{2h})$ point group, as shown in Figure 1 (a), as taken from reference \cite{Weng_PRX_2014}. Each primitive unit cell contains two ZrTe3 chains (Zr in blue and Te2 in black) linked along the c direction by two zig-zag chains formed by Te1 atoms (in red) \cite{Fje_1986, Weng_PRX_2014}. The resulting $\mathrm{ZrTe_5}$ planes are stacked along the b axis, bound by Van der Waals force, resulting in very small interlayer binding energy, comparable to the one of graphite \cite{Weng_PRX_2014}. Crystals grow naturally oriented along the \emph{a} axis exposing the \emph{ac} plane. The low dimensional nature of the prismatic chains and the weak interlayer binding energy leave sometimes the cleaved surface with multiple domains characterized by different out-of-plane chain orientations. Hence, particular attention is required during ARPES measurements (for more details see the supplementary material \cite{Suppl_mat}).


\begin{figure}[tbh]
  \includegraphics[width = 0.45 \textwidth]{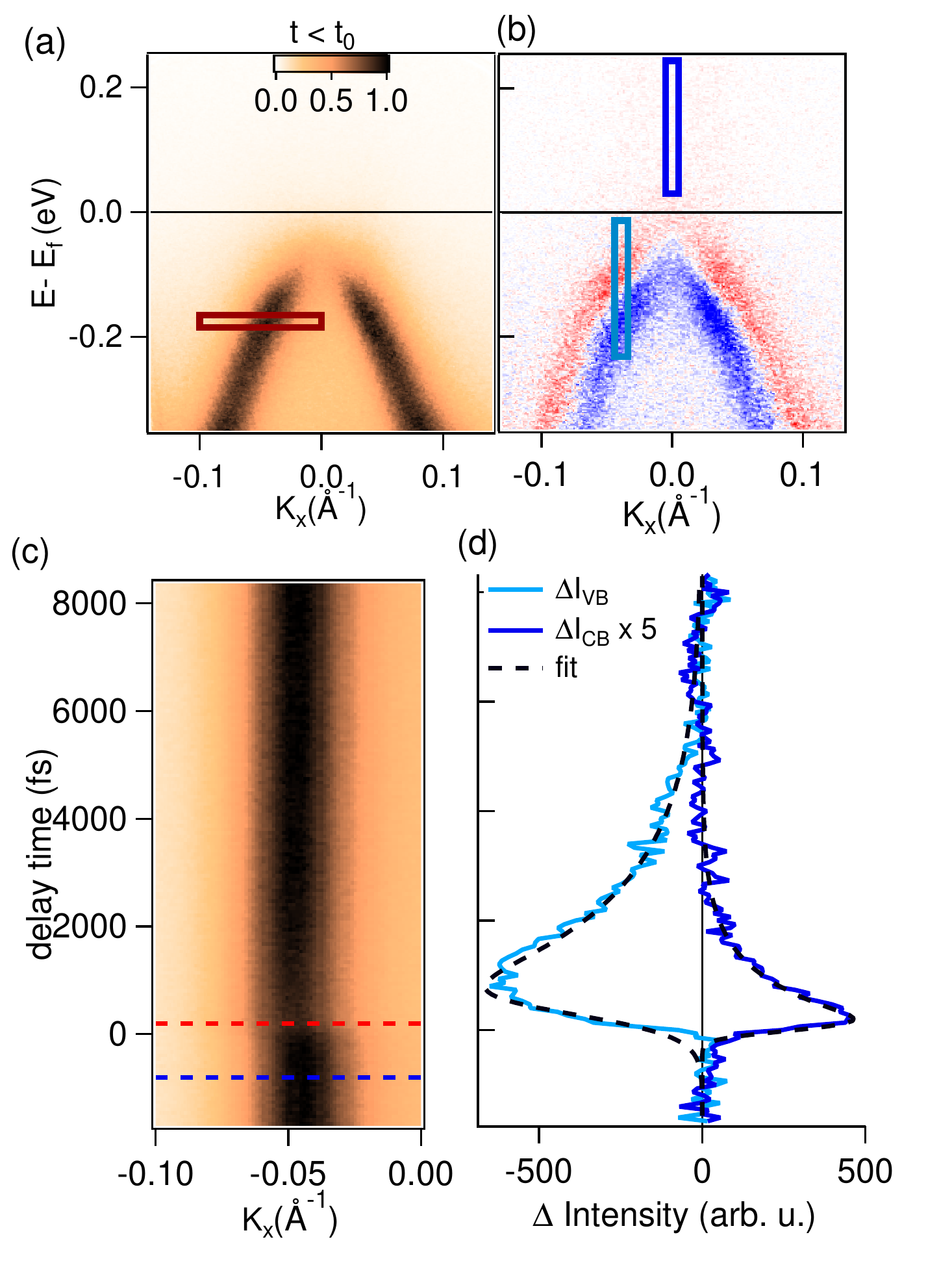}
  \caption[didascalia]{(color online)  (a) equilibrium electronic structure of $\mathrm{ZrTe_5}$ measured along the chain direction with s-polarized light at 125 K before optical excitation (-800 fs). (b) Difference image between the data acquired immediately after (200 fs) and before (-800 fs) optical excitation. The color scale provides a direct comparison with Fig. 1 (d). The close similarity suggests to interpret the results in term of a band energy shift accompanied by the increased temperature after optical perturbation. (c) Temporal evolution of the MDCs at $-0.2$ eV in the momentum window within the red rectangle of Fig. 2(a). (d) Temporal dynamics of the electron population in CB and VB, as integrated in the rectangles of Fig. 2 (b). The traces are fitted (black dashed lines) with a single decaying exponential, resulting in different characteristic times for CB, $\tau_{CB} = 0.8\pm 0.2$ ps, VB $\tau_{VB} = 1.6\pm 0.2$ ps.
    }
  \label{fig:trARPES1}
\end{figure}


We have investigated in details the temperature evolution of the electronic band structure of $\mathrm{ZrTe_5}$ along the chain direction, corresponding to the $\Gamma X$ high symmetry direction in the reciprocal space (with $\Gamma X = 0.78\AA ^{-1}$).  Owing to the low photon energy (6.2 eV) only states close to $\Gamma$ ($\pm 0.13 \AA ^{-1}$) are resolved. However previous ARPES experiments \cite{Valla_arx_2015}, band structure calculations \cite{Weng_PRX_2014} and Shubnikov �de Haas studies \cite{Levy_PRB_1985} show that the maximum of the valence band and the minimum of the conduction band are located at the $\Gamma$ point and disperse close to $\mathrm{E_F}$. Hence, only these states are responsible for the electronic transport properties. A remarkable shift of the band towards lower energies is detected upon cooling, as extracted from the analysis of the momentum distribution curves (MDCs) (for more details see the supplementary material \cite{Suppl_mat}). Results at equilibrium are summarized in Figure 1. Figure 1 (b) and (c) show the ARPES images of the valence band, dispersing with negative effective mass, measured with s-polarized light at 300 K and 125 K, respectively. The valence band (VB) is found to cross $\mathrm{E_F}$, and the Fermi wave-vector $k_F$ decreases at lower temperatures, consistently with data reported in ref. \cite{MCIllroy_JPCM_2004}.  Figure 1 (g) quantifies the magnitude of the energy shift in the investigated temperature range (300 K - 125 K), with a maximum value equal to $\sim 60 \pm 10$ meV. Figure 1 (d) shows the difference image, obtained by subtracting the band dispersion at T = 125 K from that at T = 300 K. This procedure provides a direct visualization of the energy shift and a comparison with the out-of-equilibrium tr-ARPES results, discussed later. 

In order to explain the mechanism at the origin of the $\mathrm{ZrTe_5}$ anomalous transport properties it is necessary also to access the unoccupied band structure, whose contribution becomes relevant when new bands approach $\mathrm{E_F}$ at temperature close to $\mathrm{T^{*}}$.  In order to study the band dispersion of the unoccupied conduction band (CB), two photons photoemission experiments are used. Following this scheme, the laser optical excitation transiently populates CB, which is at the same time probed by an ultrafast laser UV pulse \cite{Crepaldi_PRB_2015,Sobota_PRL_2013}. 2PPE results are interpreted as the projection of the initial state onto intermediate unoccupied states below the vacuum level. Figure 1 (e) shows the results of 2PPE experiments with p-polarized light. The intensity is almost completely suppressed in measurements with s-polarized light. Since the VB disperses around $\Gamma$ with a negative effective mass, the observed positive dispersion in ascribed to the conduction band (CB), in agreement with \emph{ab initio} calculations \cite{Weng_PRX_2014}. A linear fit of the band dispersion, dashed line, results in a band velocity of $\sim 6.8 \times 10^{5}$ m/s, which is close to $\sim 5.8 \times 10^{5}$ m/s band velocity obtained for VB. These comparable band velocities leave open the possibility to describe these states in term of linearly dispersing 3D Dirac particle,  as  recently proposed \cite{Valla_arx_2015}. Our data are compatible with an energy gap not larger than $\sim 50 \pm 10$ meV, comparable with the observed energy shift.

Figure 1(f) shows the evolution of the intensity at $\mathrm{E_F}$ as a function of temperature. MDCs are integrated in an energy window of 20 meV below  $\mathrm{E_F}$ and displayed with a vertical offset. At T = 300 K (bottom curve) we clearly distinguish the double peak structure of VB crossing $\mathrm{E_F}$. Upon cooling, the two peaks get closer and eventually merge, indicating that the top of VB lies below  $\mathrm{E_F}$. The minimum of intensity is observed at 175 K, close to the resistivity peak located at $\mathrm{T^{*} = 160}$ K. At lower temperature a new peak appears. This increase of spectral weight is attributed to the lowering of CB below $\mathrm{E_F}$. Finally, Figure 1 (g) shows, along with the energy shift, the temperature evolution of the quasiparticle lifetime $\tau_l$ extracted from the Lorentzian fit of the MDCs (for more details see the supplementary material \cite{Suppl_mat}). Between 300 K and 175 K, $\tau_l$ increases with decreasing temperature, as expected for a metal. Conversely, below $T < 175$ K,  $\tau_l$ monotonically decreases. This is regarded as signature of an increased electron scattering rate as VB is fully occupied.


Having clarified the origin of the anomalous behavior of the resistivity and the nature of the charge carriers across $\mathrm{T^{*}}$, we prove now the possibility to control the electronic properties of  $\mathrm{ZrTe_{5}}$ by mean of an ultrafast optical pulse.  Figure 2 shows the electronic properties (a) for equilibrium lattice temperature of 125 K before the arrival of the optical perturbation (-800 fs), and (b) the difference between the data immediately after (200 fs) and before (-800 fs) the arrival of the pump pulse. Figure 2 (b) displays a large energy shift of the bands very similar to that shown in Fig. 1 (c), which we interpret as follows. The optical excitation is responsible for two intertwined effects; the charge excitation from VB to CB, and the increase in the electronic energy. The latter results in the increase of both the electronic and lattice temperatures, hence accounting for the energy shift of the band structure. However, few fs-temporal resolution would be necessary to disentangle the electronic contribution from the lattice one. Figure 2 (c) shows the evolution of the MDCs at $\mathrm{E - E_{F}} = -0.2$ eV for $-0.1\AA^{-1}<k<0\AA^{-1}$ in the region highlighted with the rectangle in Fig.2 (a). This small momentum window is chosen in order to better visualize the dynamics of the band shift as a consequence of the transient increase in temperature.

One of the key results reported in this Letter is the capability to manipulate, at the ultrashort timescale, the electronic band structure of $\mathrm{ZrTe_5}$. At the same time, we also report on the dynamics of the charge transfer between VB and CB. Figure 2 (d) shows the temporal evolution of the electron population in VB (negative signal) and CB (positive signal) integrated in the energy-momentum windows enclosed by rectangles in Fig. 2 (b). The integration windows are chosen in order to account in the same momentum window for the energy shift of the band. A single decaying exponential fit of the CB dynamics gives a characteristic time $\tau_{CB} = 0.8\pm 0.2$ ps, which is significantly different from $\tau_{VB} = 1.6\pm 0.2$ ps measured for VB. We notice also that the maximum excitation in VB is delayed with respect to CB. These findings indicate that different scattering mechanisms are responsible for both the electron (hole) accumulation at the bottom (top) of CB (VB) and the subsequent relaxation. 


\begin{figure}[t]
 \includegraphics[width = 0.5 \textwidth]{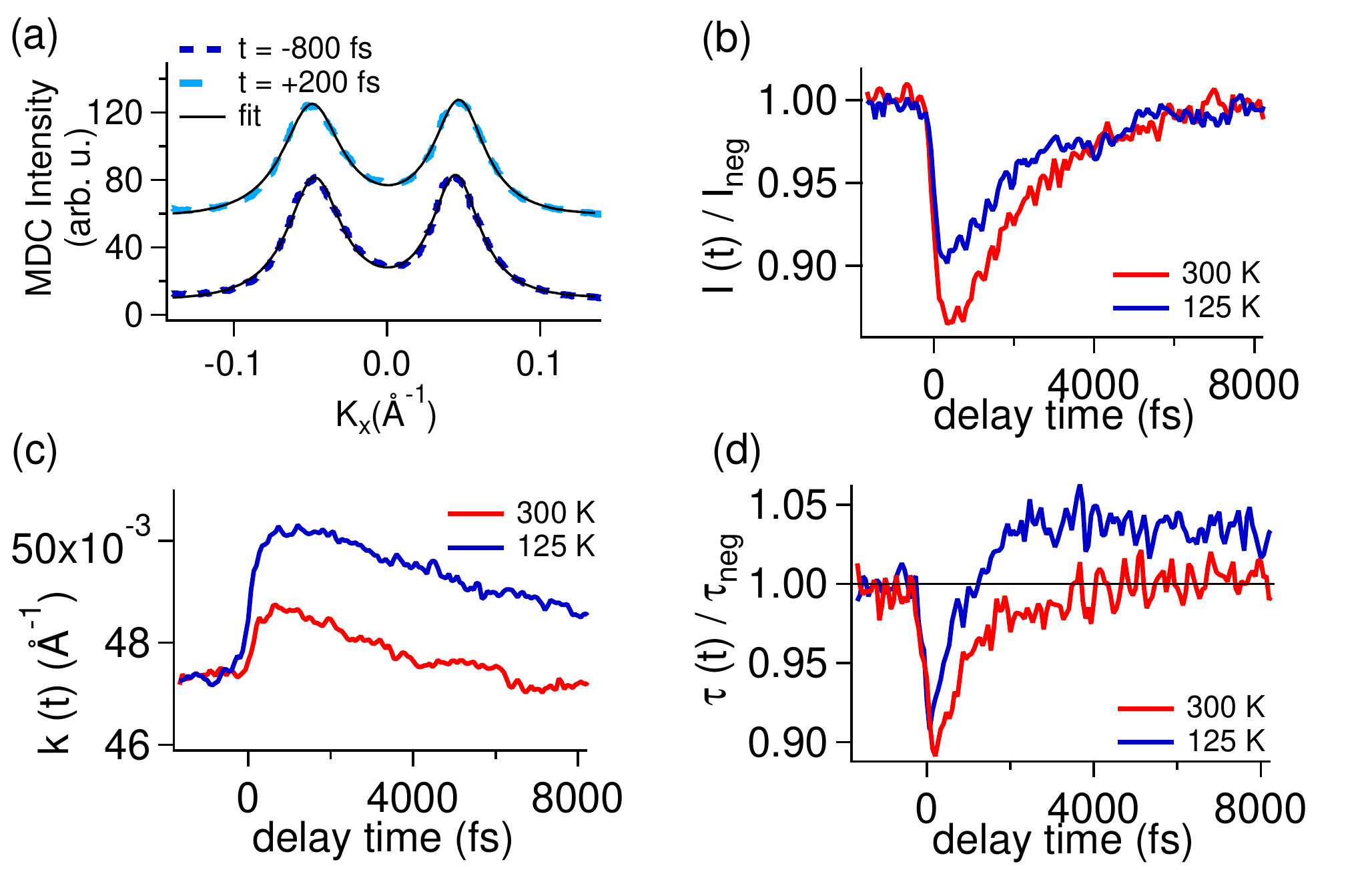}
  \caption[didascalia]{(color online) (a) MDCs extracted for a lattice temperature at equilibrium of 125 K at -0.2 eV, as measured before (-800 fs, dark blue) and after (200 fs, light blue) optical excitation, displayed with a vertical offset. The best fit curves (black lines) are obtained with a double Lorentzian function plus a polynomial background. (b-d) results of the analysis of the Lorentzian fit to the MDCs as extracted at -0.2 eV (blue, 125 K) and -0.14 eV (red, 300 K). The two energies are chosen in order to compare the band peak dynamics at the same \emph{k} position. (b) dynamics of the left Lorentzian peak intensity at 300 K and 125 K, normalized to the negative time values. (c) Comparison between the peak positions,  reflecting the band energy shift. (d) temporal evolution of the quasiparticle lifetime $\tau_l$, a significant increase in lifetime is observed for long delay time, only for the equilibrium lattice temperature of 125 K.
  
   }
  \label{fig:trARPES_2}
\end{figure}


Figure 3 summarizes the dynamics of the optically induced band energy shift. We compare the temporal dynamics of the MDCs measured at equilibrium lattice temperature of 125 K (blue) and 300 K (red). For these two temperatures, the MDCs are extracted at -0.2 eV and -0.14 eV, respectively, in order to compare the dynamics of the emission peak at the same \emph{k} value. Figure 3 (a) shows, with a vertical offset, two MDCs for equilibrium lattice temperature of 125 K before (-800 fs, dark blue) and after optical excitation (200 fs, light blue), along with the best fit (black line). Each MDC is fitted with a double Lorentzian function plus a polynomial background, as for the analysis in Fig.1. From each fit we obtain the temporal evolution of the Lorentzian function parameters: intensity, position of the maximum and width. Hereafter, we limit the discussion to the results for the band branch at negative \emph{k} values. Figure 3 (b) shows the evolution of the intensity, normalized to the value at equilibrium. The intensity decreases after optical excitation as a consequence of the combined increase in temperature (broadening of the Fermi Dirac distribution) and the charge excitation from VB to CB. Figure 3 (c) shows the dynamics of the maximum intensity position, which shifts to larger momentum values reflecting the band energy shift. 

Finally, we focus on the Lorentzian width, which is expressed in term of the quasiparticle lifetime $\tau_l$, normalized to the value before optical excitation. In stark contrast to intensity and maximum intensity position, which have qualitatively the same dynamics both for T = 300 K and 125 K, $\tau_l$ shows different behaviors for the two different equilibrium lattice temperatures. At short time scale, both at 300 K and 125 K, $\tau_l$ shortens after optical excitation. While at 300 K, $\tau_l$ recovers its equilibrium value with a single decaying exponential, with characteristic time $1.1$ ps, at 125 K $\tau_l$ changes sign and increases by $\sim 5 \%$ of the equilibrium value. The observed increase in the quasiparticle lifetime is consistent with the increase in the electronic and lattice temperature, as shown in Fig. 1 (g). 


In conclusion, in this Letter we report a comprehensive investigation of both the occupied (VB) and unoccupied (CB) states of $\mathrm{ZrTe_5}$ revealing a binding energy shift of these bands as a function of temperature. This remarkable finding unveils the origin of the resistivity anomaly at $\mathrm{T^*} \sim 160$ K along with the charge carriers switch from holes to electrons upon decreasing the temperature across $\mathrm{T^*}$.
Having clarified the mechanism at the origin of the transport anomaly of $\mathrm{ZrTe_5}$, we have experimentally proven the possibility to optical modify the electronic properties of this material both in term of binding energy and quasiparticle lifetime of the valence band, thus leading to the control, at the ultrafast time scale, of the transport properties of $\mathrm{ZrTe_5}$. These results provide an external knob to control the $\mathrm{ZrTe_5}$ conductivity and to unlock the route for a unique platform for magneto, optical and thermoelectric transport applications.

Very recently we became aware that a similar effect of temperature dependent energy shift of the electronic band structure has been reported by ARPES experiments on $\mathrm{WTe_2}$ \cite{Kaminski_Arxiv_2015}, a material which also displays unique electronic properties \cite{Ali_Nat_2014}.  This suggests that the mechanism studied in this Letter is general and of fundamental importance in controlling the transport properties of a potentially wide class of materials

This work was supported in part by the Italian Ministry of University and Research under Grant Nos. FIRBRBAP045JF2 and FIRB-RBAP06AWK3 and by the European Community Research Infrastructure Action under the FP6 "Structuring the European Research Area" Program through the Integrated Infrastructure Initiative "Integrating Activity on Synchrotron and Free Electron Laser Science", Contract No. RII3-CT-2004-506008. We acknowledge support by the Swiss NSF.





\end{document}